\documentclass[conference]{IEEEtran}
\usepackage{url}
\usepackage{ upgreek }
\usepackage{verbatim}
\usepackage{amsfonts,amssymb}
\usepackage{threeparttable}
\usepackage{algorithmic}
\usepackage{array}
\usepackage{verbatim}

\usepackage{color}
\usepackage{mathtools}
\newtheorem{lemma}{Lemma}
\usepackage{ dsfont }
\usepackage[misc]{ifsym}
\usepackage[utf8]{inputenc}
\usepackage{cite}

\usepackage{algorithm} 
\usepackage{algorithmic}  
\usepackage[algo2e]{algorithm2e} 

\ifCLASSINFOpdf
   \usepackage[pdftex]{graphicx}
\else
\fi
\usepackage{amsmath}
\usepackage{array}
\usepackage{algorithm}
\usepackage{url}

\usepackage{graphicx}
\usepackage{subfigure}  

\begin{document}
\title{Storage Control for Carbon Emission Reduction: Opportunities and Challenges}

\author{\IEEEauthorblockN{Jian Sun, Yaoyu Zhang, Yang Yu \emph{and} Chenye Wu$^\text{\Letter}$}
\IEEEauthorblockA{Institute for Interdisciplinary Information Sciences\\
Tsinghua University, Beijing, 100084, P.R. China\\
Email:chenyewu@tsinghua.edu.cn}
}

\maketitle

\begin{abstract}
Storage is vital to power systems as it provides the urgently needed flexibility to the system. Meanwhile, it can contribute more than flexibility. In this paper, we study the possibility of utilizing storage system for carbon emission reduction. The opportunity arises due to the pending implementation of carbon tax throughout the world. Without the right incentive, most system operators have to dispatch the generators according to the merit order of the fuel costs, without any control for carbon emissions. However, we submit that storage may provide necessary flexibility in carbon emission reduction even without carbon tax. We identify the non-convex structure to conduct storage control for this task and propose an easy to implement dynamic programming algorithm to investigate the value of storage in carbon emission reduction.

\vspace{0.1cm}
\end{abstract}


%
\IEEEpeerreviewmaketitle

\section{Introduction}

Global warming is coming \cite{1}. However,  implementing carbon tax is still under debate, less to say an agreement on the value of carbon tax. Hence, in the near future, the power system operator will still have to stick to its conventional way to conduct economic dispatch: collect the bids from generators (which mostly reflect the generators' fuel costs), and select the fuel cost effective generators to meet the demand, although many fuel cost effective generators (e.g., coal fire power plants) are producing huge amount of carbon emissions. 

\subsection{Alternative to Reduce Carbon Emission}
Besides carbon tax, we identify an alternative to reduce carbon emission: to smartly use storage systems for emission reduction. The idea is that when conducting economic dispatch, although the system operator has to dispatch the generators according to the merit order in fuel cost, it can choose to use storage systems to avoid the dispatch of  some carbon intensive generators. However, this is a delicate task in that different from conducting economic dispatch for minimizing the social cost, our proposed way of carbon emission reduction leads to a \emph{non-convex} optimization problem. The \emph{non-convexity}  precisely comes from the fixed dispatch order for the system operator (i.e., the merit order in fuel cost).

\begin{figure}
    \centering
    \includegraphics[scale=0.25]{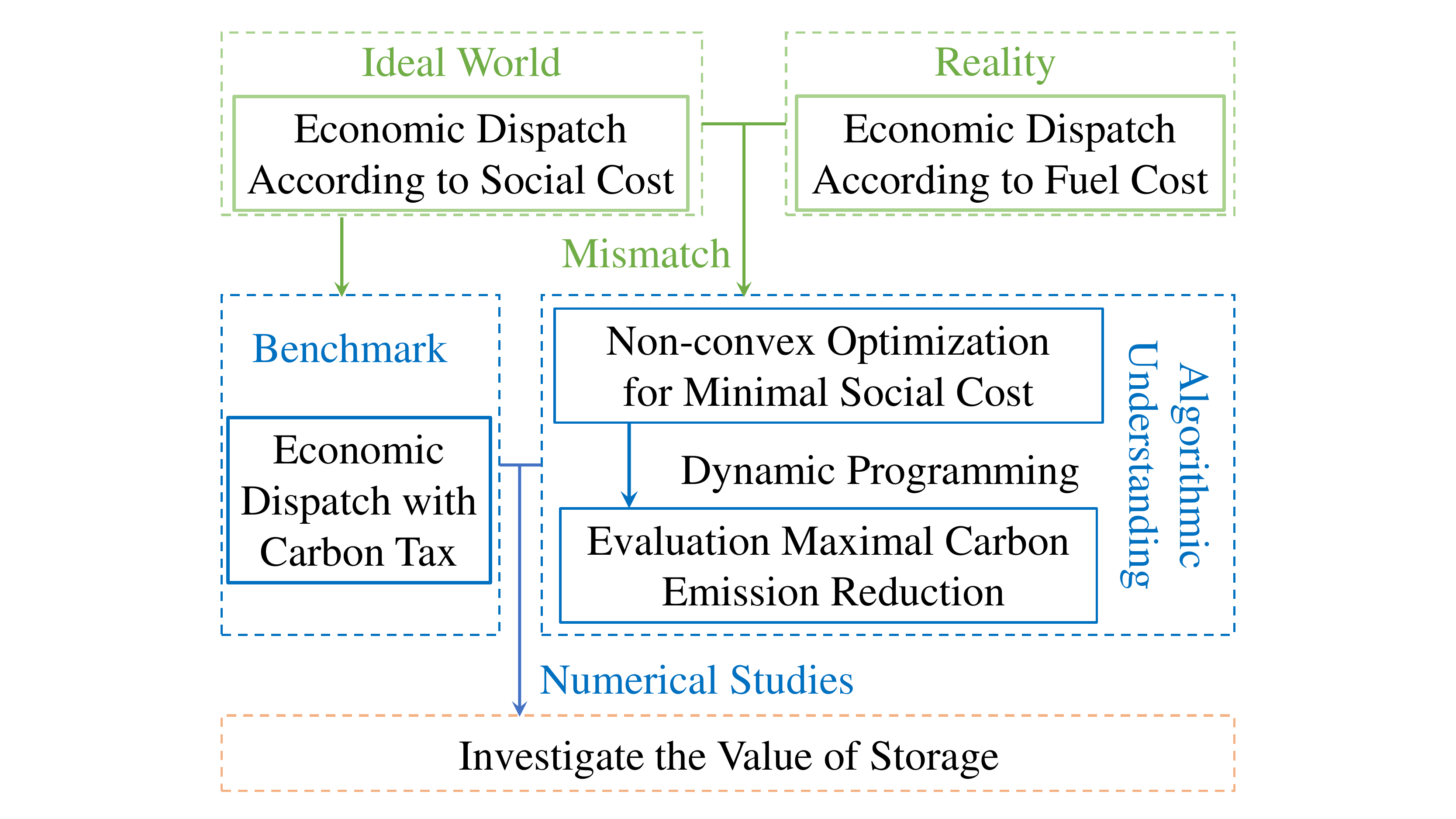}
    \caption{Our paradigm to investigate the value of storage.}\vspace{-0.5cm}
    \label{structure}
\end{figure}

In this paper, we seek to exploit the structure of the non-convex optimization problem, and propose an easy to implement algorithm, which serves as the basis for us to examine the value of storage in reducing carbon emissions. Figure \ref{structure} visualizes this paradigm.

\subsection{Related Works}

The major body of related literature focuses on utilizing storage systems for power system control and electricity market operation. Just to name a few, Vojvodic \emph{et al.} present a multistage stochastic programming framework to improve the hydroelectric plants' efficiency by utilizing the pumped storage in \cite{EJOR}. Nottrott \emph{et al.} propose a linear programming model for the photovoltaic battery storage system to determine the optimal storage dispatch schedules in \cite{2}. Wu \emph{et al.} investigate the sharing economy business model for storage sharing in an industrial park in \cite{4}. 
Koutsopoulos \emph{et al.} seek to minimize long-term average grid operational cost by deriving a near optimal threshold-based storage control policy in \cite{19}. 
    
While the research on policy design for carbon emission reduction has caught much attention (see  \cite{17} and \cite{18} for more detailed comparisons between carbon tax and its variants), the literature rarely considers utilizing storage system for carbon emission reduction. Zöphel \emph{et al.} investigate the marginal value of installed storage system for the German energy system in terms of different CO$_2$ prices in \cite{10}.

\subsection{Our Contributions}
Different from the literature, we provide an algorithmic treatment. More specifically, we use dynamic programming (DP) to solve the non-convex optimization problem for evaluating the value of storage. The principal contributions of this paper can be summarized as follows:

\begin{enumerate}
    \item \emph{Non-convexity Characterization:} Using mathematical formulation as well as numerical examples, we highlight the non-convexity in examining the value of storage in reducing carbon emissions. 
    \item \emph{Algorithm Design:} To tackle the difficulty induced by non-convexity, we propose a practical DP algorithm to minimize the social cost using storage systems. 
    \item \emph{Performance Assessment:} We prove the approximation bound for the discretization in DP, and use numerical studies to highlight the trade-off between running time (efficiency) and accuracy. With carefully selected parameters, we investigate the value of storage in carbon emission reduction via simulation.
\end{enumerate}

The rest of the paper is organized as follows. Section \ref{system model} introduces system model and highlights the non-convexity in the optimization problem. We propose DP algorithm to solve the non-convexity in Section \ref{section 3}. Section \ref{simulation_sec} studies the efficiency and accuracy of our proposed scheme and investigates the value of storage. Finally, concluding remarks and future directions are given in Section \ref{conclusion_sec}.

\section{System Model}
\label{system model}
To highlight the non-convexity in the decision making, we consider an electricity pool model, and ignore the transmission line constraints. In the pool model, the system operator will simply dispatch the generators based on the merit order of their fuel costs.

\subsection{Formulation of System Cost}

We denote $\text{MF}_i$ as the fuel cost of generator $i$. According to system operator' economic dispatch, we formulate the cost minimizing optimization to meet this demand of $x$. Based on this formulation, we can then define the total fuel cost $f(x)$ to meet this demand of $x$ as follows:
\begin{align*}
    f(x) \coloneqq \min\nolimits_{g_{i}} &\sum\nolimits_{i=1}^n \text{MF}_i\cdot g_i\\
     s.t. &\sum\nolimits_{i=1}^n g_i=x\\
     &0\le g_i \le \Bar{g}_i
   \tag{1}
\end{align*}
where $g_i$ denotes generator $i$'s output, $\Bar{g}_i$ denotes $g_i$'s capacity limit, and $n$ is the total number of generators in system.

\begin{figure}[t]
    \centering
    \includegraphics[scale=0.25]{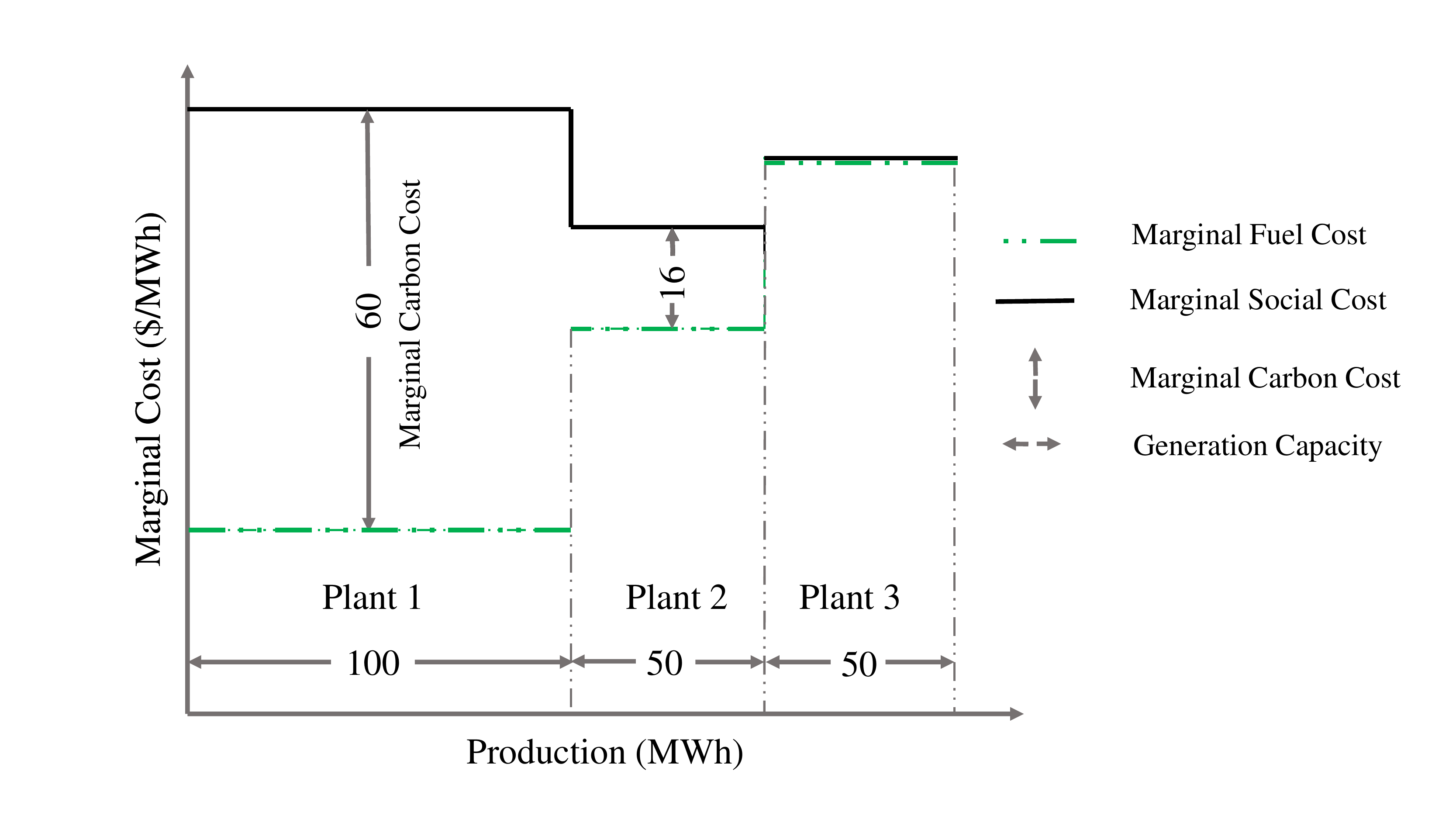}
    \caption{Example to highlight the non-convexity in $C'(x)$.}\vspace{-0.5cm}
    \label{marginal cost}
\end{figure}

Note that the demand $x$ is a parameter of economic dispatch which yields the total fuel cost fuction $f(x)$. As the demand $x$ increases from 0 to $\sum_{i=1}^n \Bar{g}_i$, the dispatch order is the merit order in fuel cost. Denote this dispatch order by $\pi=(\pi(1),...,\pi(n))$, which is a permutation of the $n$ generators. According to this dispatch order, we can define the carbon cost to meet demand of $x$ given $\pi$:
\begin{align*}
    c(x)  \coloneqq \min_{k,g_{\pi(k)}} &\text{MC}_{\pi(k)}\cdot g_{\pi(k)}\!+\!\!\sum\nolimits_{i=1}^{k-1} \text{MC}_{\pi(i)}\cdot \Bar{g}_{_{\pi(i)}}\\
     s.t. & \sum\nolimits_{i=1}^{k-1} \Bar{g}_{_{\pi(i)}}\le x < \sum\nolimits_{i=1}^{k} \Bar{g}_{_{\pi(i)}}\\
    & g_{\pi(k)}=x-\sum\nolimits_{i=1}^{k-1} \Bar{g}_{_{\pi(i)}}
    \tag{2}
\end{align*}
where $\text{MC}_k$ denotes the marginal carbon cost of generator $k$. Based on the definitions of $f(x)$ and $c(x)$, we can define the total social cost to meet demand of $x$ given $\pi$ as follows:
\begin{equation*}
    C(x)=f(x)+c(x) \tag{3}
\end{equation*}
While the marginal fuel cost ($f'(x)$) is non-decreasing, the marginal social cost ($C'(x)$) may not. This is because the dispatch order $\pi$ is determined by minimizing the total fuel cost. Figure \ref{marginal cost} visualizes a three generator example to highlight the non-convexity in the marginal social cost. In this example, the marginal fuel costs of the three generators are 30, 60, and 80$\$$/MWh, respectively. And the merit order of the three generators is exactly $\pi=(1,2,3)$. However, they correspond with different carbon costs. Hence, if the system operator has to dispatch the generators according to $\pi$, then the marginal social cost $C'(x)$ is non-convex (as shown in the solid black line in Fig. \ref{marginal cost}).

\subsection{Formulation for Storage System}
Given the fixed dispatch order, one way to reduce the total social cost is to utilize the storage systems. In this paper, we assume the system operator is equipped with a storage device of capacity $B$. Denote the state of charge in the storage at the time $t$ by $s_t$, and the system demand at time $t$ by $D_t$. Then, at each time $t$, with the help of storage, the system operator should dispatch the generators and acquire energy of $x_t$ from the grid at time $t$:
\begin{equation*}
    x_t=s_t + D_t -s_{t-1} \tag{4}
\end{equation*}
\subsection{Cost Minimization with Storage Control}
The system operator may seek to minimize the expected social cost with the storage system over time span [1,...,$T$]. The resulting optimization problem can be cast as follows:
\begin{align*}
    \min_{s_1\ldots s_t} &\quad \mathbb{E}\left\{\sum\nolimits_{t=1}^T C(s_t + D_t -s_{t-1})\right\} & \\
    s.t. &\quad 0\le x_t \le X \coloneqq \sum\nolimits_{i=1}^n\Bar{g}_i &\quad \forall t\\
    &\quad 0\le s_t \le B &\quad \forall t
    \label{objective}
    \tag{5}
\end{align*}
Note that in this formulation, we highlight the fact that demand $D_t$ is a random variable. The randomness comes from the prediction error. One may choose to consider the renewable generation as negative load. Then the randomness comes from the prediction error for both load and renewable generation. The problem, though very much simplified from the reality, is still challenging to solve due to the non-convexity in the total social cost function $C(x)$.

\section{Algorithm: Dynamic Programming}
\label{section 3}
In this section, we seek to exploit the structure of the non-convex optimization problem, and propose an easy-to-implement DP algorithm for the solution. 

\subsection{Recursive Structure}
To better investigate the value of storage in reducing carbon emissions, we require $s_0=s_T=\frac{1}{2} B$. This excludes storage's opportunity of arbitraging, and also achieves the maximal flexibility in storage control. However, $s_T=\frac{1}{2}B$ serves as the boundary condition in DP formulation. For a general decision making problem, we define $OPT(\tau,s,B)$ as follows:
\begin{align*}
        \min_{s_1...s_{\tau}} &\quad \mathbb{E} \left \{\sum\nolimits_{t=1}^{\tau} C(s_t+D_t-s_{t-1})\right \}&\\
    s.t.& \quad 0\le x_t\le X &\forall t\\
    &\quad s_0=\frac{1}{2} B\\
    &\quad s_{\tau}=s\\
    & \quad 0\le s_t\le B &\forall t
    \tag{6}
\end{align*}

For $OPT(\tau,s,B)$, we can define the optimal control actions by $S_{\tau}=\{s_1\ldots s_{\tau}\}$ and define $\mu_{\tau}(s)$ as the corresponding minimal social cost. Then by the linearity of expectation, we can conclude that 
\begin{equation*}
    \mu_\tau(s)\! =\! \min_{s'\in R^s_{\tau\!-\!1}}\!\big \{\mathbb{E}\!\{\!C(s\!-\!s'\!+\!D_{\tau})\!\}\!+\!\mu_{\tau\!-\!1}(s')\!\big \}
    \tag{7}
\end{equation*}
where $R_{\tau-1}^s$ is the feasible region for $s'$. This recursive structure holds for any set of parameters to obtain the optimal storage control strategies.

\subsection{Dynamic Programming}
While the true feasible region for any $s'$ is a continuous space, this poses the challenges for implementing the DP due to curse of dimensionality. However, in our setting, since the cost function is continuous and bounded, we can devide the feasible region of $s'$ into $\delta$ steps (each step of size $\delta$), and conduct the DP on the discretized feasible region. We submit that, this will only incur a bounded optimization error. More precisely, we can prove the following Lemma:
\begin{lemma}
The $\delta$-step discretization for problem $OPT(\tau,s,B)$ will incur an optimization error of at most $\bar{M}_c\cdot\tau\cdot\delta$, where $\bar{M}_c$ is the maximal derivative of $C(\cdot)$.
\label{lemma}
\end{lemma}
We provide the proof in the Appendix.Based on the $\delta$-step discretization, we illustrate our algorithm below:

\begin{algorithm}
    \caption{Recursive $[S_{\tau},\mu_{\tau}]=OPT(\tau,k\delta,B)$}
    \label{alg}
\begin{algorithmic}
\REQUIRE 
    $\tau$: number of hours; \\
$k\delta$: $k\delta$ is the target state of charge by hour $\tau$;\\
$B$: the battery capacity
\ENSURE
$S_{[\tau]}=\{s_1,...,s_{\tau}\}$: optimal storage control actions;\\
$\mu_{\tau}$:the corresponding minimal social cost\\

    \IF{$\tau==1$}
        \STATE return \quad $S_{1}=\{s_1\}=\{k\delta\}$\
        \STATE \qquad \qquad $\mu_{1}=\mathbb{E}\{C(k\delta-s_0+D_1)\}$\
    \ELSE
    \FOR{$i\delta \in R_{\tau-1}^s$}
\STATE $opt[i]=OPT(\tau\!-\!1,i\delta,B)\!+\!\mathbb{E}\{C(k\delta\!-\!i\delta\!+\!D_{\tau})\}$\
\ENDFOR
    \ENDIF
    \STATE $i^*=\text{argmin}$ $opt[i]$\
    \STATE $[S^*_{\tau-1},\mu^*_{\tau-1}]=OPT(\tau-1,i^*\delta,B)$\
    \STATE return \quad $S_{\tau}=S^*_{\tau-1}\cup \{s_{\tau}=k\delta\}$\
    \STATE \qquad \qquad $\mu_{\tau}=\mu^*_{\tau-1}+\mathbb{E}\{C(k\delta-i^*\delta+D_\tau)\}$\
\end{algorithmic}
\end{algorithm}

Note that in the worst case, we need to enumerate $B/\delta$ states  in each decision stage. Hence, the space complexity is $O(TB/\delta)$ and the time complexity is $O(TB^2/\delta^2)$.

\vspace{0.2cm}
\noindent \textbf{Remark}: We want to emphasize that the space and time complexity make the DP look like a polynomial algorithm. However, since the input size is $\log_2B$, instead of $B$, it still suffers from curse of dimensionality for an accurate solution. Nonetheless, with the bound of approximation error, we can select a reasonable $\delta$ to relieve the computational burden. Also, DP is easy to implement by nature. Hence, we argue our proposed algorithm is still quite practical.

Another often neglected challenge in the DP formulation is the calculation of $\mathbb{E}\{C(k\delta-i\delta+D_{\tau})\}$ given the distribution of $D_{\tau}$ (or the load prediction error distribution). The naive approach to using $C(k\delta-i\delta+\mathbb{E}(D_{\tau}))$ to estimate $\mathbb{E}\{C(k\delta-i\delta+D_{\tau})\}$ is not accurate due to the non-convexity in the cost function. One solution is to use look-up-table for an efficient query, as the demand distribution is periodic by nature. Nonetheless, in the numerical study, we assume a prefect prediction for the demand in the near future time span $[1,...,T]$ to highlight the value of storage.

\begin{figure}[t]
   \centering
    \subfigure[]{\includegraphics[width=1.7in]{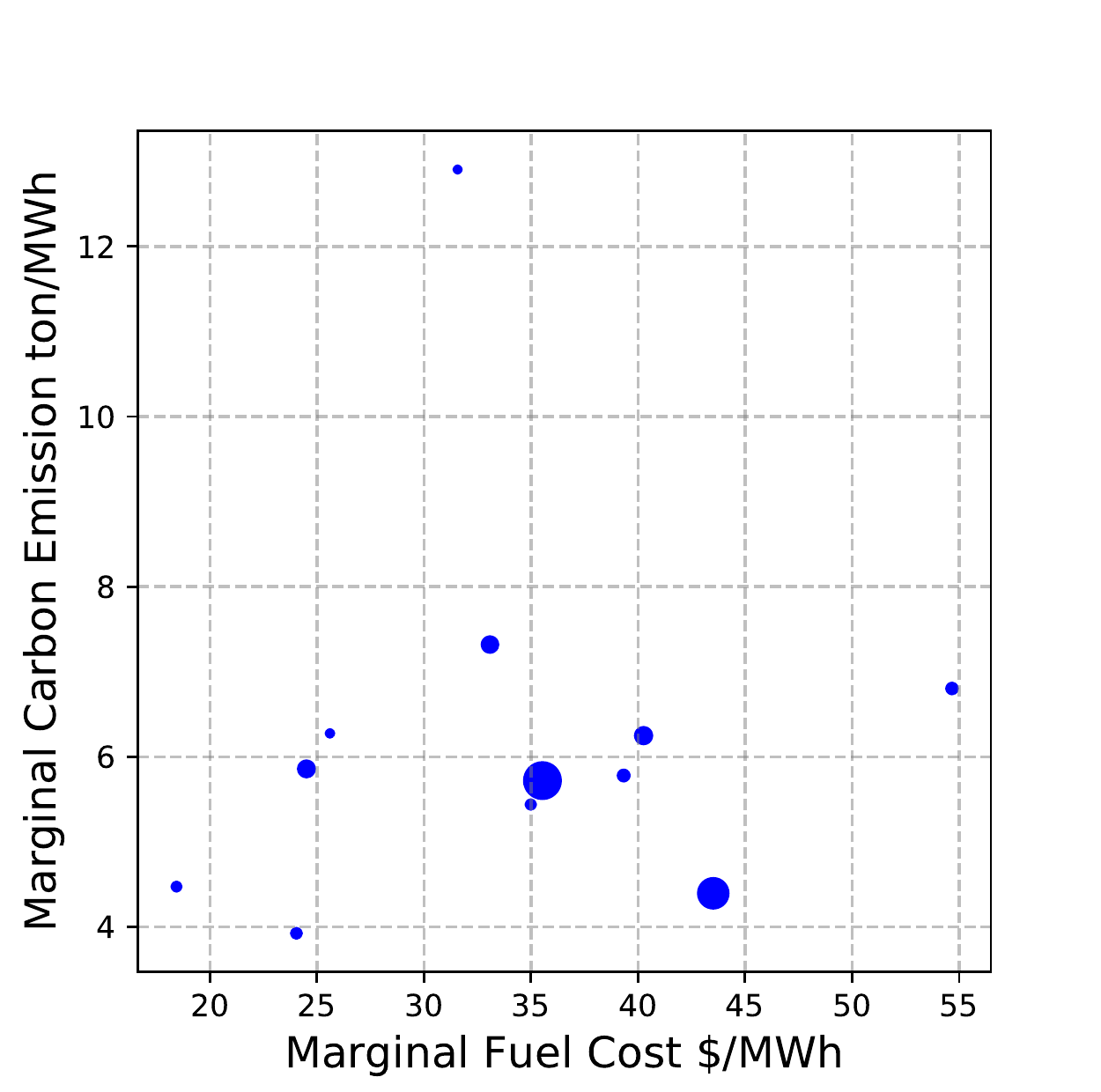}}
    \subfigure[]{\includegraphics[width=1.7in]{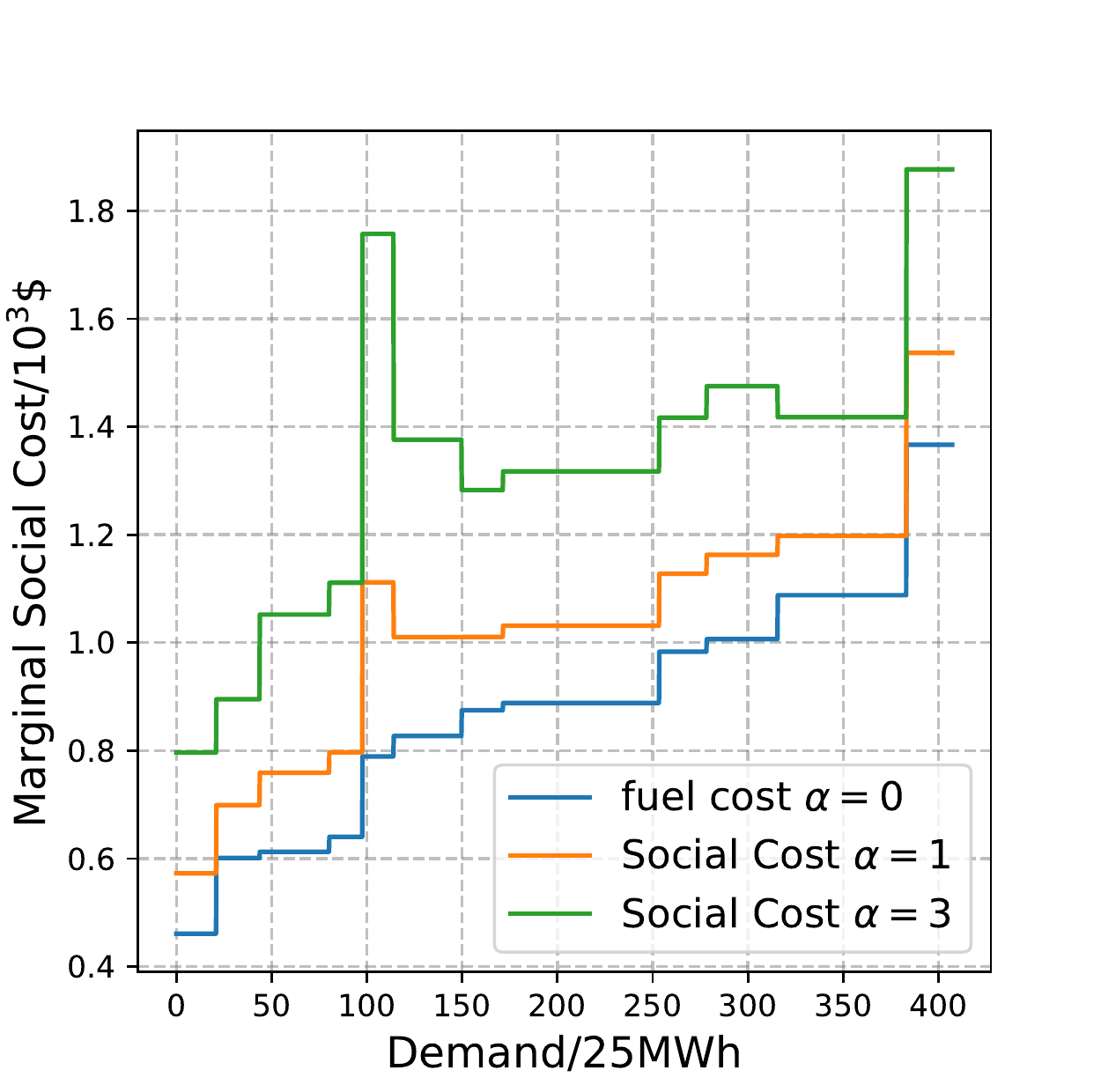}}
    \caption{Statistical features of the generators}
    \label{Statistical features of the generators}
 \end{figure}
 
 \begin{figure}[t]
    \centering
    \includegraphics[scale=.38]{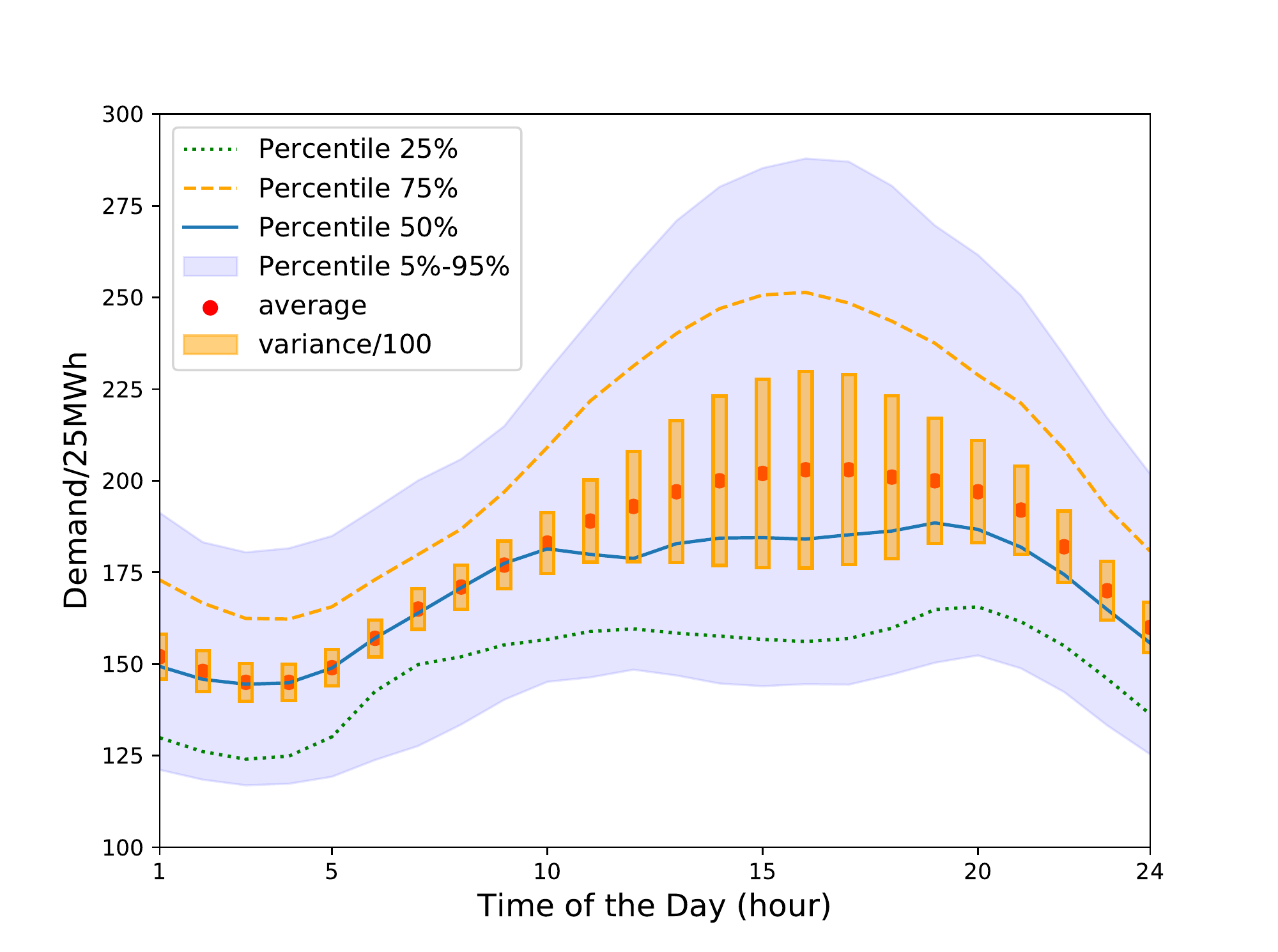}
    \caption{Statistical features of demand}
    \label{demand distribution}
\end{figure}

\begin{figure}[t]
    \centering
    \includegraphics[scale=.38]{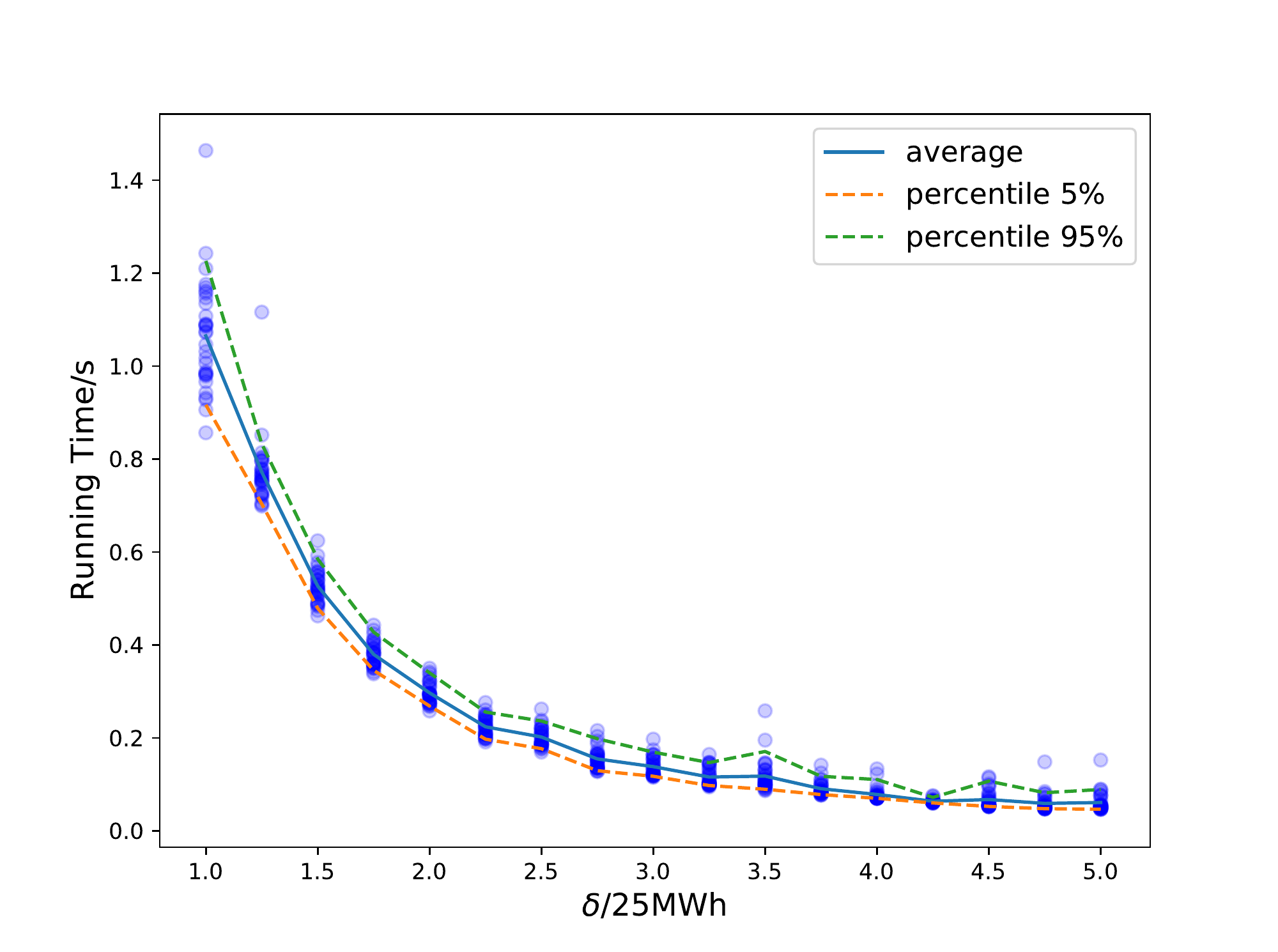}
    \caption{Running time with $\delta$-discretization}
    \label{Running Time}
\end{figure}

\begin{figure}[t]
    \centering
    \includegraphics[scale=.38]{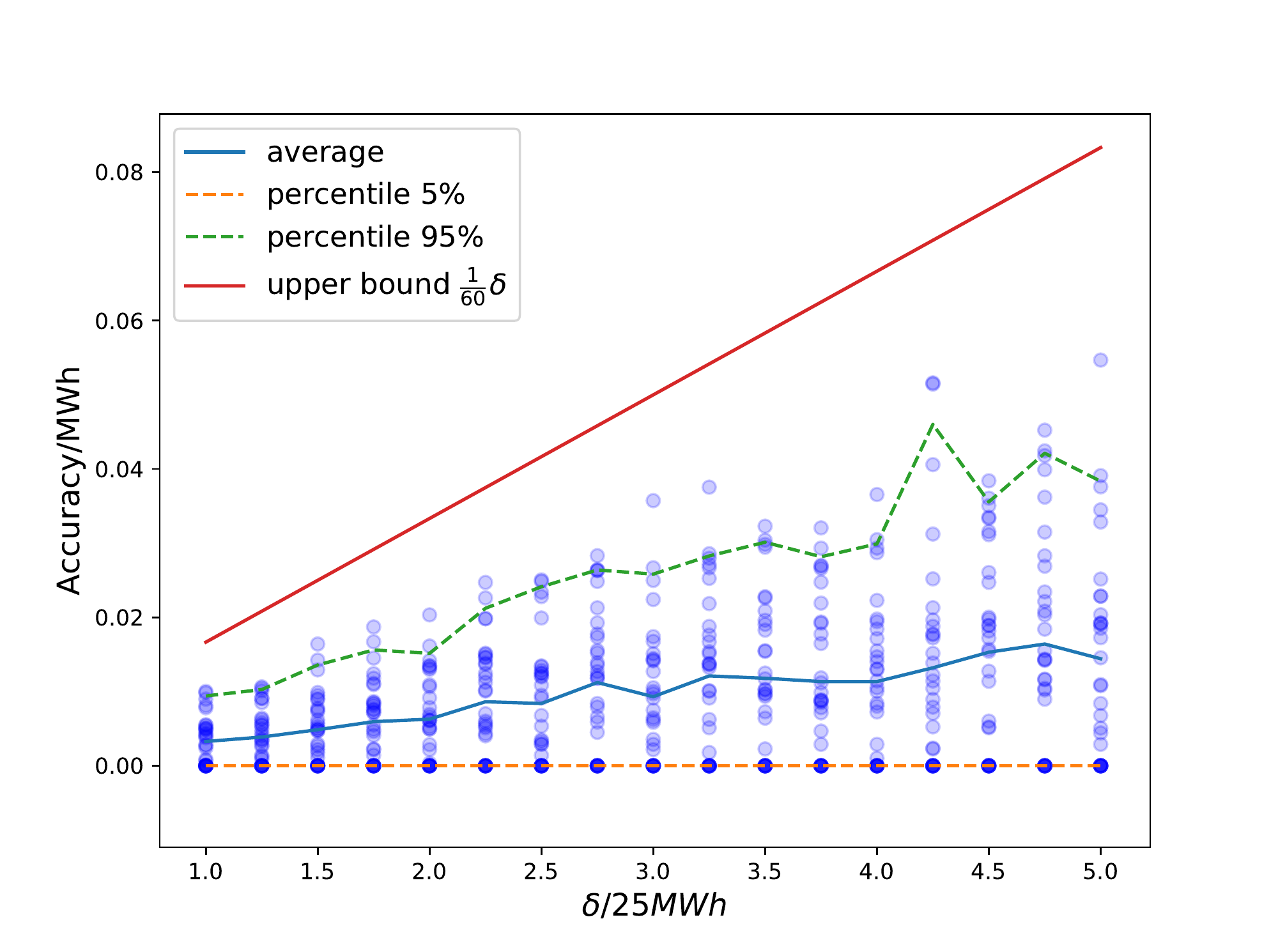}
    \caption{Accuracy with $\delta$-discretization}
    \label{Accuracy}
\end{figure}

\begin{figure*}[t]
    \centering
    \subfigure[Social cost reduction]{\includegraphics[width=2.2in]{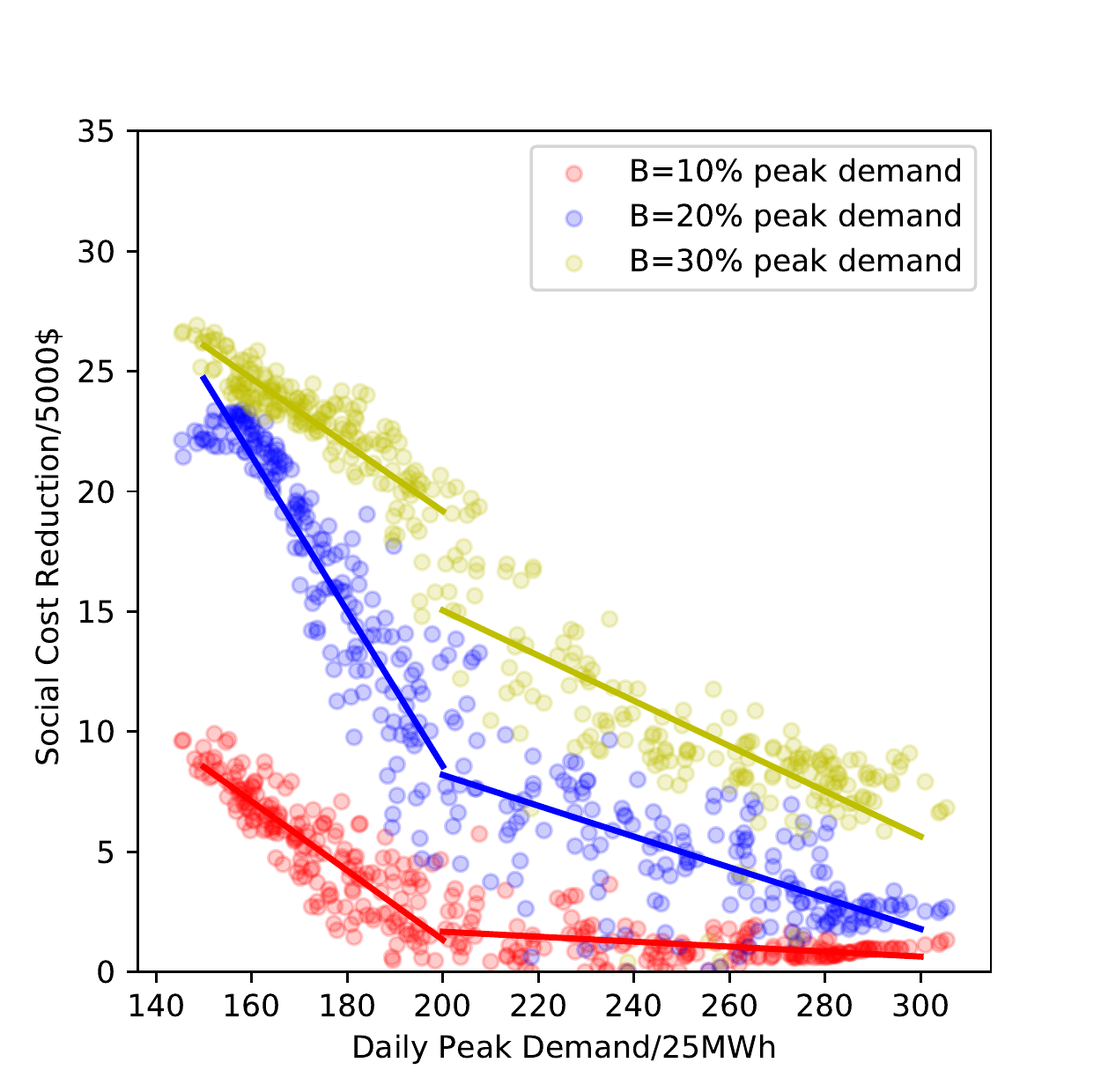}}
    \subfigure[Carbon cost reduction]{\includegraphics[width=2.2in]{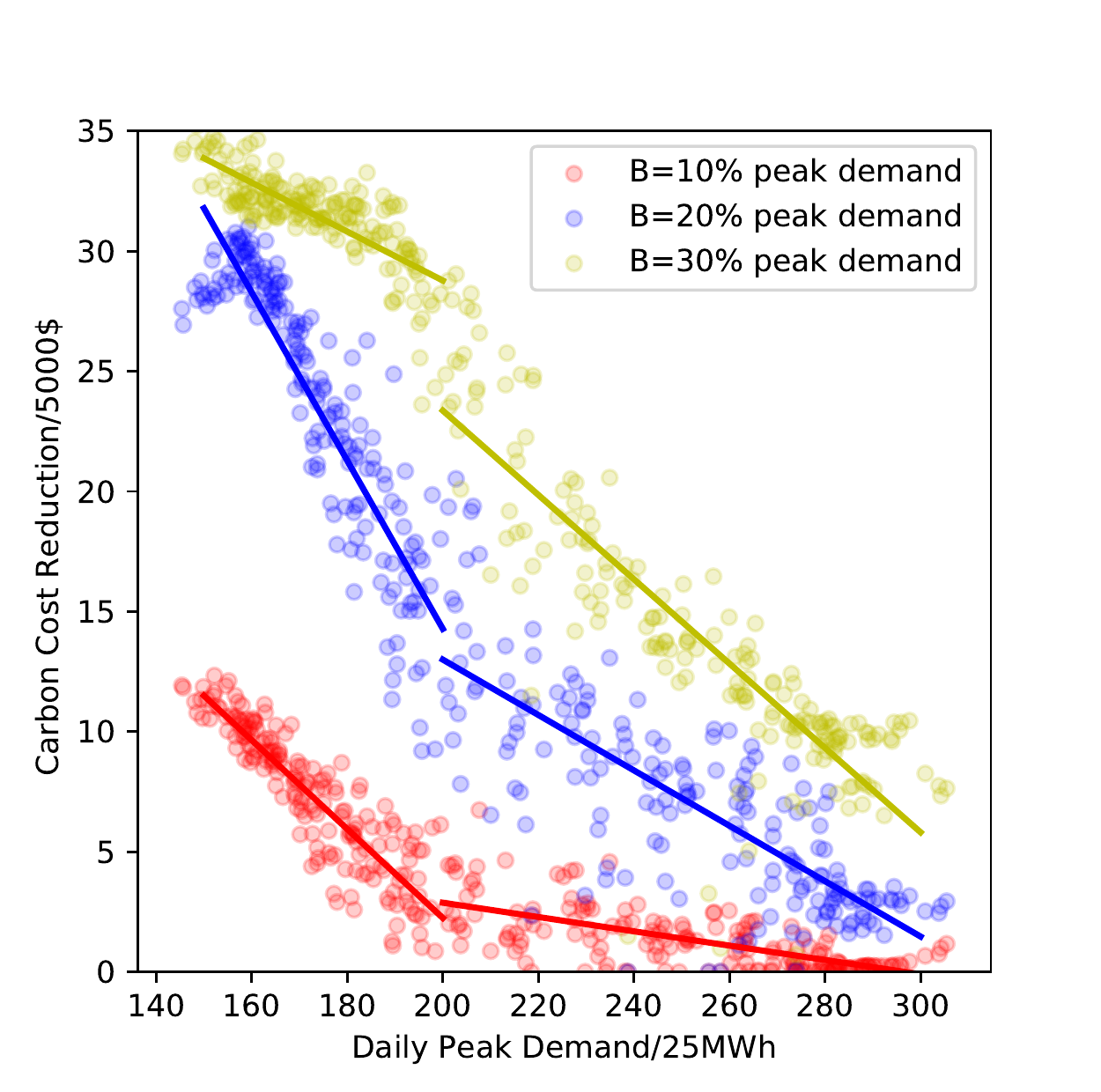}}
    \subfigure[Fuel cost reduction]{\includegraphics[width=2.2in]{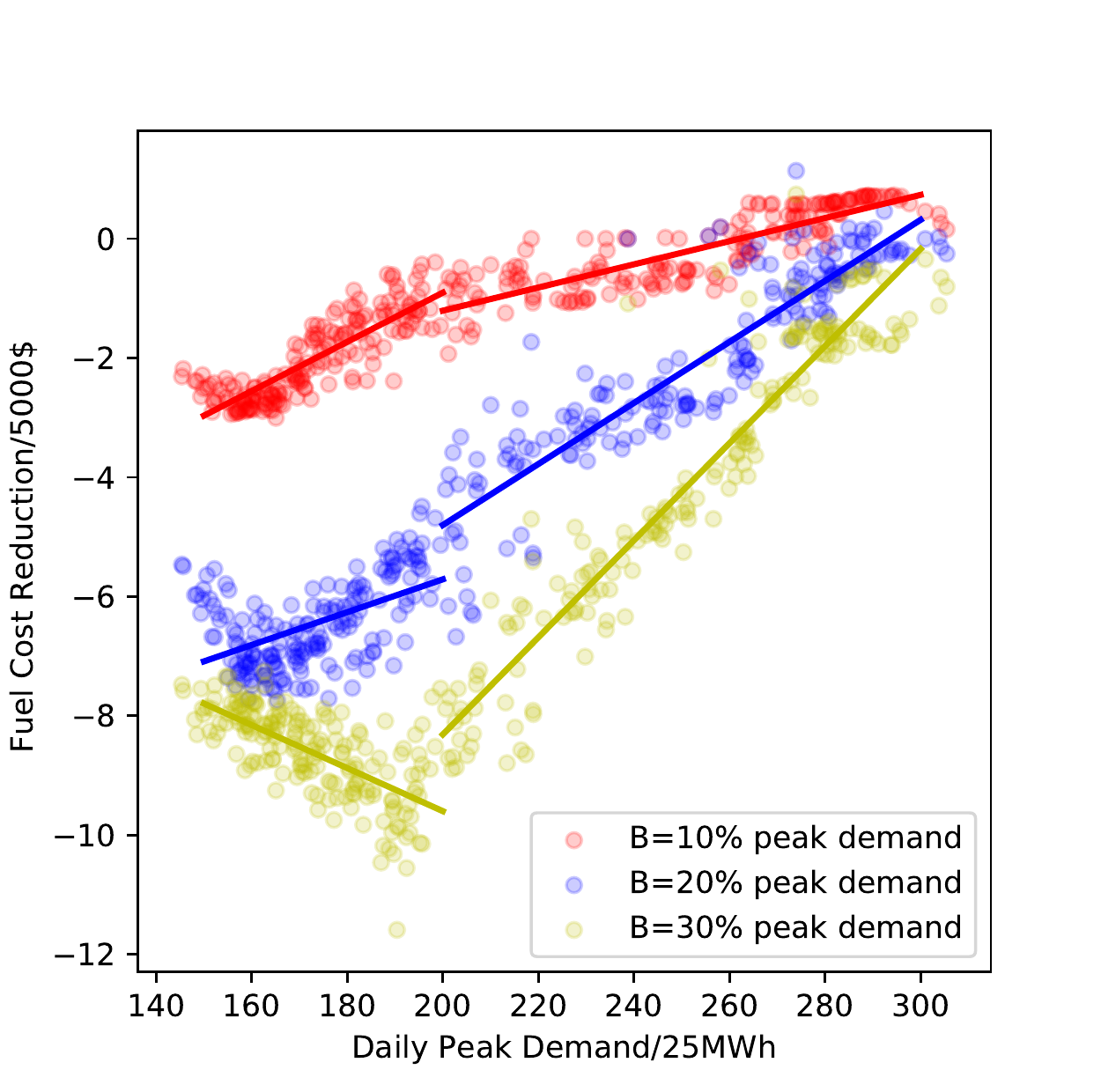}}
    \caption{Cost Reduction when $\alpha=3$}\vspace{-0.3cm}
    \label{Cost Reduction}
\end{figure*}

\begin{figure}[t]
    \centering
    \includegraphics[width=2.6in]{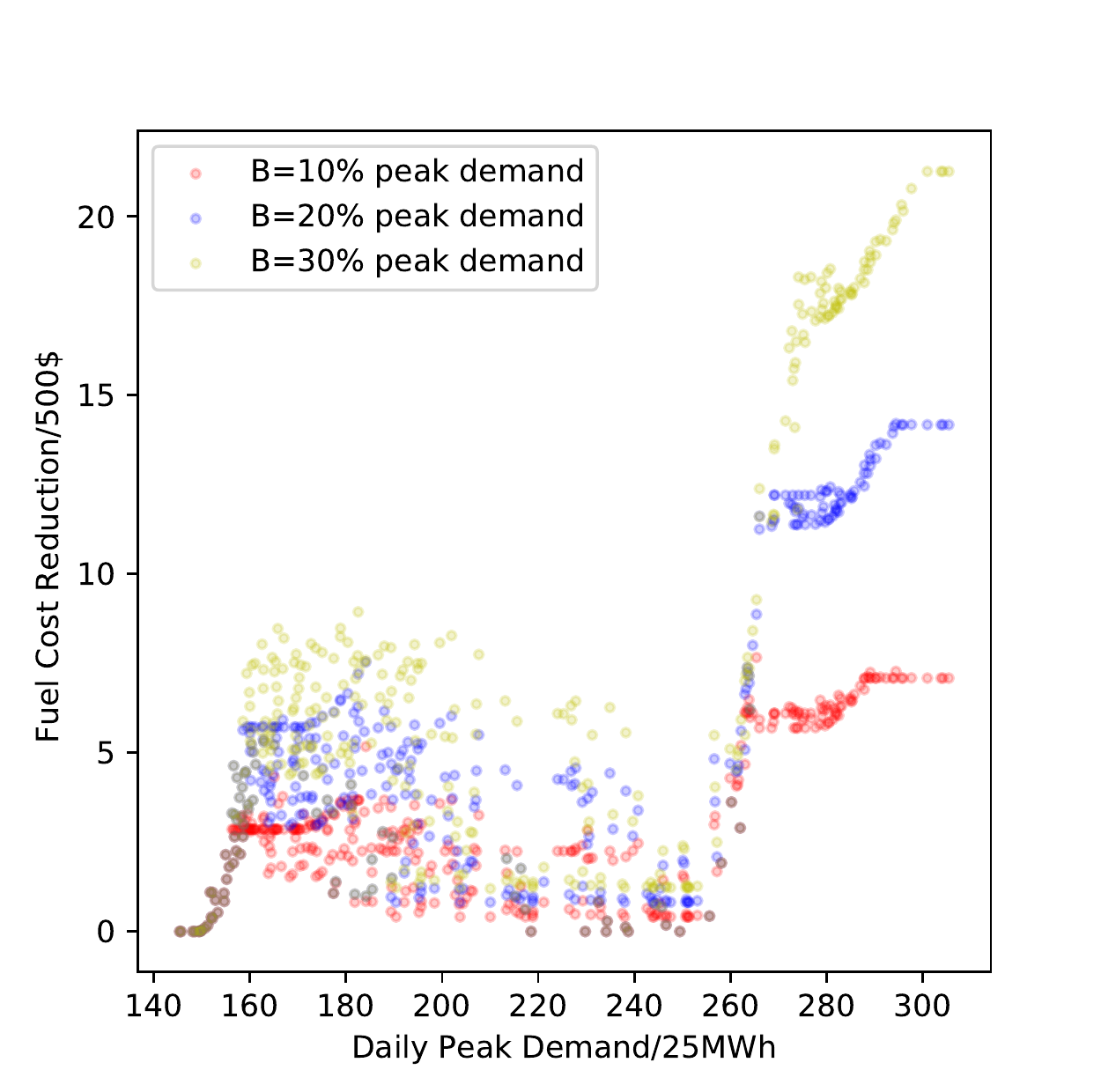}
    \caption{Fuel cost reduction without carbon cost}
    \label{Fuel Cost Reduction}
\end{figure}

\section{Simulation}
\label{simulation_sec}
To evaluate the efficiency and effectiveness of our proposed framework, we combine the data from
EIA (US energy information Administration)\cite{23} and ERCOT (Electric reliability council of Texas)\cite{24} to obtain the true generator parameters (including generation capacity, fuel cost and carbon emissions). Figure. \ref{Statistical features of the generators}(a) plots the generator statistics (larger circle implies larger capacity). \par
Note that, the observed data is carbon emission, instead of carbon cost ($\text{MC}_i$ in our context). To relate the carbon emission (denoted by $\text{MCE}_i$ for generator $i$) and the carbon cost, we define a parameter $\alpha$, such that 
\begin{equation*}
    \text{MC}_i=\alpha \cdot \text{MCE}_i, \quad \forall i
    \tag{8}
\end{equation*}

One way to explain the physical nature of $\alpha$ is the carbon tax. Figure \ref{Statistical features of the generators}(b) plots the marginal social cost corresponding to various $\alpha$, which again highlights the non-convexity in $C(\cdot)$.

We use the annual demand data from ERCOT in the year of 2018 and scale the annual peak demand be smaller than the maximal capacity in the system. We plot the statistical features of the annual demand profiles in Fig.\ref{demand distribution}.

\subsection{Efficiency Evaluation}
We evaluate our DP algorithm's efficiency by observing the trend of its running time with the increasing of step size $\delta$. In the simulation, we assume the storage capacity is 20$\%$ of the peak demand. Figure \ref{Running Time} verifies that the algorithm running time can be reduced significantly with a slight longer $\delta$. On the other hand, Fig. \ref{Accuracy} demonstrates the algorithm accuracy varying with $\delta$. Here, we define accuracy $\gamma$ as follows:
$$\gamma={\lim_{\delta_0 \to 0}}\dfrac{C(\delta)-C(\delta_0)}{\bar{M}_c\cdot\tau}$$
where $C(\delta)$ denotes the minimal social cost with $\delta$-step discretization, and the denominator is for better comparison of our derived optimization bound in Lemma \ref{lemma}.

In Fig. \ref{Accuracy}, the solid red line is $\frac{1}{60}\delta$, which far more smaller than the theoretical bound (which is $\delta$). This implies our proposed DP algorithm performs reasonably well in practice.

\subsection{Effectiveness Evaluation}
To investigate the value of storage in reducing carbon emission, we select $\alpha$ to be 3 for the simulation. Figure. \ref{Cost Reduction} plots the daily social cost, carbon cost and fuel cost reductions over a year. As shown in Fig. \ref{Cost Reduction}, our proposed scheme achieves the maximal social cost reduction and carbon cost reduction when the daily peak demand is relatively small. And the reduction rates decrease as the daily peak demand increases. This illustrates that when the peek demand is low, there are more flexibility in reducing the carbon emissions. However, the reduction comes from a higher fuel cost as shown in Fig. \ref{Cost Reduction}(c). This highlights the non-convexity in the social cost function.

We also compare the social cost reduction with the case when there is no carbon cost ($\alpha=0$). Through pure storage arbitraging, the fuel cost reduction shows a non-monotone pattern. Note that the unit in Fig. \ref{Cost Reduction} is per 5000\$ while the unit in Fig. \ref{Fuel Cost Reduction} is per 500\$. Hence, in terms of cost reduction, our scheme easily outperforms the storage arbitraging in most range of daily peak demand. However, this also raises the concern that the incentives to conduct storage control are not aligned with arbitraging, which warrants the design of side payments. However, a detailed discussion is beyond the scope of our paper.

\section{Conclusion}
\label{conclusion_sec}
In this paper, we seek to understand the value of storage in carbon emission reduction. By identifying the non-convex structure in the decision making problem, we design a practical DP algorithm using $\delta$-step discretization with performance guarantee. We examine the value of storage via numerical studies on the carbon cost reduction.

This work can be extended in many ways. As we have mentioned in Section \ref{section 3}, it is interesting to further explore the structure of the problem and make the DP more efficient. It is also important to understand the role of network constrains when exploiting the value of storage. In this work, we haven't rigorously characterized the randomness in the storage control problem. However, one important feature of storage system is its capability of providing flexibility to tackle the challenges induced by the uncertainties in the system. We plan to use stochastic programming to investigate the value of storage system under uncertainties.

\bibliographystyle{ieeetr}
\bibliography{reference}

\begin{thebibliography}{10}

\bibitem{1}
Y.~Xu, V.~Ramanathan, and D.~G. Victor, ``Global warming will happen faster
  than we think,'' {\em Nature}, vol.~564, no.~7734, pp.~30--32, 2018.

\bibitem{EJOR}
G.~Vojvodic, A.~I. Jarrah, and D.~P. Morton, ``Forward thresholds for operation
  of pumped-storage stations in the real-time energy market,'' {\em European
  Journal of Operational Research}, vol.~254, no.~1, pp.~253--268, 2016.

\bibitem{2}
A.~Nottrott, J.~Kleissl, and B.~Washom, ``Energy dispatch schedule optimization
  and cost benefit analysis forgrid-connected, photovoltaic-battery storage
  systems,'' {\em Renewable Energy}, vol.~55, no.~4, pp.~230--240, 2013.

\bibitem{4}
C.~Wu, D.~Kalathil, K.~Poolla, and P.~Varaiya, ``Sharing electricity storage,''
  in {\em Decision \& Control}, 2016.

\bibitem{19}
I.~Koutsopoulos, V.~Hatzi, and L.~Tassiulas, ``Optimal energy storage control
  policies for the smart power grid,'' in {\em IEEE International Conference on
  Smart Grid Communications}, 2011.

\bibitem{17}
G.~I. Galinato and J.~K. Yoder, ``An integrated tax-subsidy policy for carbon
  emission reduction,'' {\em Resource \& Energy Economics}, vol.~32, no.~3,
  pp.~310--326, 2010.

\bibitem{18}
D.~Pearce, ``The role of carbon taxes in adjusting to global warming,'' {\em
  Economic Journal}, vol.~101, no.~407, pp.~938--948, 1991.

\bibitem{10}
C.~Zophel and D.~Most, ``The value of energy storages under uncertain
  co2-prices and renewable shares,'' in {\em European Energy Market}, 2017.

\bibitem{23}
EIA, ``Energy information administration.'' \url{https://www.eia.gov/}.

\bibitem{24}
ERCOT, ``Electric reliability council of texas.'' \url{http://ercot.com.}

\end{thebibliography}

\section*{Appendix: Proof for Lemma 1}
\noindent \emph{Proof:} Denote the derivative of $\mu_{\tau}(s)$ by $\dot \mu_{\tau}(s)$. Then the key to prove Lemma \ref{lemma} is to identify the following Fact:

\textbf{Fact 1:} $\dot \mu_{\tau}(s) \le \Bar{M}_c$

To prove the Lemma, it suffices to show
\begin{equation*}
    \big|\mu_i(s)\!-\!\tilde{\mu}_i(s)\big|\le \bar{M}_c\cdot\tau\cdot\delta
    \tag{9}
    \label{by fact}
\end{equation*}
where $\Tilde{\mu}_{\tau}(s)$ is the output of the DP algorithm with $\delta$-step discretization. $\Tilde{\mu}_\tau(s)\! =\! \min_{k\delta\in R^s_{\tau\!-\!1}}\!\big \{\mathbb{E}\!\{\!C(s\!-\!s'\!+\!D_{\tau})\!\}\!+\!\Tilde{\mu}_{\tau\!-\!1}(s')\!\big \}$ This can be proved by induction.

The induction basis is when $\tau=1$: Eq. (\ref{by fact}) trivially holds.

Assume Eq. (\ref{by fact}) holds for $\tau=i-1$. Denote $\bar{\mu}_\tau(s)\! =\! \min_{k\delta\in R^s_{\tau\!-\!1}}\!\big \{\mathbb{E}\!\{\!C(s\!-\!k\delta\!+\!D_{\tau})\!\}\!+\!\mu_{\tau\!-\!1}(k\delta)\!\big \}$, Using Fact 1, the following holds:
\begin{equation}
    \big|\mu_i(s)\!-\!\bar{\mu}_i(s)\big|\le \bar{M}_c\cdot\delta.
    \tag{10}
\end{equation}
Hence, together with the induction hypothesis and Eq. (10), when $\tau=i$,
\begin{align*}
    \big|\mu_i(s)-\Tilde{\mu}_i(s)\big|\le& \big|\mu_i(s)-\bar{\mu}_i(s)\big|+\big|\bar{\mu}_i(s)-\Tilde{\mu}_i(s)\big|\\
    \le& \Bar{M}_c\cdot \delta+\bar{M}_c\cdot (i-1)\cdot\delta=\bar{M}_c\cdot i\cdot\delta
\end{align*}
which concludes the proof.$\hfill\blacksquare$

\end{document}